\documentclass{article}
\usepackage[dvips]{graphicx}
\usepackage{fullpage}

\begin{document}


\newcommand{\ada}{adiabatic approximation}
\setlength{\unitlength}{1cm}
\newcommand\bra[1]{\left\langle#1\right|}
\newcommand\ket[1]{\left|#1\right\rangle}
\newcommand\eqref[1]{(\ref{#1})}
\newcommand\beq{\begin{equation}}
\newcommand\eeq{\end{equation}}
\newcommand\bea{\begin{eqnarray}}
\newcommand\eea{\end{eqnarray}}
\newcommand\eps{\epsilon}
\newcommand\ltwid{\mathrel{
 \raise.3ex\hbox{$<$\kern-.75em\lower1ex\hbox{$\sim$}}}}
\newcommand\la{\lambda}
\newcommand\ga{{\gamma}}

\begin{flushright}
\vskip-0.75cm
UdeM-GPP-TH-05-140\\
\end{flushright}
\bigskip

\begin{center}
{\LARGE A new perturbative approach to the adiabatic approximation}\\
\bigskip\bigskip
R. MacKenzie,$^{1,2}$ E. Marcotte$^1$ and H. Paquette$^1$\\
\it $^1$\ Physique des particules, Universit\'e de Montr\'eal\\
C.P. 6128, Succ. Centre-ville, Montr\'eal, QC H3C 3J7\\
$^2$\ D\'epartement IRO, Universit\'e de Montr\'eal\\
C.P. 6128, Succ. Centre-ville, Montr\'eal, QC H3C 3J7\\
\end{center}

\bigskip
\begin{abstract}
A new and intuitive perturbative approach to time-dependent quantum mechanics problems is presented, which is useful in situations where the evolution of the Hamiltonian is slow. The state of a system which starts in an instantaneous eigenstate of the initial Hamiltonian is written as a power series which has a straightforward diagrammatic representation. Each term of the series corresponds to a sequence of ``adiabatic" evolutions, during which the system remains in an instantaneous eigenstate of the Hamiltonian, punctuated by transitions from one state to another. The first term of this series is the standard adiabatic evolution, the next is the well-known first correction to it, and subsequent terms can be written down essentially by inspection. Although the final result is perhaps not terribly surprising, it seems to be not widely known, and the interpretation is new, as far as we know. Application of the method to the adiabatic approximation is given, and some discussion of the validity of this approximation is presented.
\end{abstract}
\bigskip


The {\ada} in quantum mechanics was developed in the very early days of quantum mechanics \cite{Born:1928}, and is presented in many of the classic textbooks on the subject (for one of the most detailed presentations, see Messiah \cite{Messiah:1962}). The essential idea is very simple. A time-dependent Hamiltonian has instantaneous eigenstates and eigenenergies, which are the solutions to the time-independent Schroedinger equation using $H(t)$, where $t$ is viewed as a parameter rather than the time. We can visualize the evolution of the system in terms of a graph of the instantaneous energies and eigenstates as a function of time. Suppose the system is initially in an instantaneous eigenstate of the Hamiltonian. Then if the Hamiltonian evolves slowly enough, the system will remain in the state which evolves from the initial state. However, exactly what constitutes ``slowly enough" remains the subject of some debate.

In the 1980s, the {\ada} came to the forefront again with the work of Berry
\cite{Berry:1984jv}, who showed that the wave function of a slowly-evolving system acquires a phase factor previously thought to be of no physical significance. Berry's phase has been observed in a wide variety of physical contexts, and has been proposed in many more; see the book by Shapere and Wilczek \cite{Shapere:1989kp} for a detailed discussion and some of the early experimental and theoretical investigations.

More recently, the advent of the field of quantum computing has revived interest in the
{\ada} \cite{Marzlin:2004, Sarandy:2004, Pati:2004, Tong:2004, Ambainis:2004, Wu:2004, Tong:2005, Ye:2005}. One reason for this is a paradigm of quantum computation based on the evolution of the ground state of a Hamiltonian designed to interpolate adiabatically between a simple Hamiltonian with known ground state and another Hamiltonian from whose ground state the answer to an interesting problem can be found \cite{Farhi:1996na, Farhi:2000}.

A second reason is simply that in many approaches to quantum computing qubits are manipulated by changing an external influence (such as a magnetic field on a spin), hopefully resulting in adiabatic evolution of the qubits.

In both these situations, it is obviously important to understand the conditions under which the evolution of the system can be considered adiabatic.

In this paper, we present a new perturbative approach to time-dependent quantum mechanics problems, which is useful in the adiabatic limit. The state of the system is written as a power series in the number of jumps undertaken by the system from one instantaneous eigenstate of the Hamiltonian to another; these jumps are connected by an adiabatic evolution in the instantaneous state in which the system finds itself between jumps. To do this, we divide the time evolution into infinitesimal increments. Within one of these time slices the Hamiltonian can be treated perturbatively and a perturbative expansion for the evolution within each time slice can be calculated. The resulting series can be visualized as a diagrammatic expansion which is extremely intuitive and with which higher-order terms can be written down by inspection.

The first term of the series gives the usual adiabatic expression for the final state; the second is the standard correction to the adiabatic evolution (see, for example, Messiah \cite{Messiah:1962}).

A somewhat simpler derivation of the final result can in fact be given by an iterative solution of the integral version of the time-dependent Schroedinger equation (writing the state as a sum over the instantaneous eigenstates of the Hamiltonian with coefficients to be determined), but in our opinion the derivation presented below is sufficiently elegant and intuitive that it is worth the slight additional complexity.

Our result can be compared to other investigations of corrections to the {\ada}. Berry \cite{Berry:1987a} has calculated the phase change acquired by the state in ``super-adiabatic" time evolution the physical interpretation, thus determining corrections to the Berry phase, as an iterative sequence of approximations to the phase.  Moody, Shapere and Wilczek \cite{Moody:1989vh} have gone beyond the {\ada}, calculating perturbatively the probability of transition to other states, resulting in a sort of depletion of the adiabatic state. Below, we will see that our result can be shown to agree with these. Our method, however, is quite different, and the physical interpretation, as far as we know, is new.

Consider  a time-dependent Hamiltonian $H(t)$. We define the instantaneous eigenstates and eigenenergies $\ket{n(t)}$ and $E_n(t)$ as the solutions to the time-independent Schroedinger equation (parameterized by time):
\beq
H(t)\ket{n(t)}=E_n(t)\ket{n(t)}.
\label{one}
\eeq
For simplicity, we assume that there are no degeneracies throughout the time evolution. The state $\ket{n(t)}$ is said to have evolved from $\ket{n(0)}$.

According to the {\ada}, if the Hamiltonian evolves ``slowly enough", a system prepared initially in an eigenstate of $H(0)$ will, at time $T$, be in the state which evolved from the initial state. Let us see how this happens.

If the initial state is $\ket{\psi(0)} =\ket{0(0)}$ (not necessarily the instantaneous ground state of the system), then
\beq
\ket{\psi(T)} =U(T)\ket{0(0)},
\label{onea}
\eeq
where the time evolution operator is
\beq
U(T)=\hat T\exp -i\int_0^T dt \,H(t),
\label{two}
\eeq
in units where $\hbar=1$  and where $\hat T$ denotes time ordering.

We divide the total time interval $T$ into $N$ intervals of duration $\eps=T/N$; ultimately we will take the limit $N\to\infty$, $\eps\to0$ with $T$ fixed. Define $t_j=j\eps$, the beginning of the $j$th time interval. Then we can write
\beq
U(T)=U_{N-1}U_{N-2}\cdots U_{1}U_{0},
\label{three}
\eeq
where $U_j$ is the evolution operator taking the system from $t_j$ to $t_{j+1}$:
\beq
U_j=\hat T\exp -i\int_{t_j}^{t_{j+1}} dt \,H(t).
\label{four}
\eeq

Let us find an expression for the state after one time interval, $\ket{\psi(t_1)}=U_0\ket{\psi(0)}=U_0\ket{0(0)}$. It is useful to expand this state in eigenstates of $H(t_1)$:
\bea
\ket{\psi(t_1)}&=&\sum_m\ket{m(t_1)}\bra{m(t_1)}U_0\ket{0(0)}\nonumber\\
&=&\sum_m\ket{m(t_1)}\bra{m(t_1)}\hat T\exp -i\int_0^{t_1} dt \,H(t)\ket{0(0)}.
\label{five}
\eea
So far, no approximation has been made. We will now do so. Writing
$H(t)=\sum_n\ket{n(t)}E_n(t)\bra{n(t)}$, we expand each of these factors about $t=0$:
\bea
\ket{n(t)}&=&\ket{n(0)}+t\ket{\dot n(0)}+\cdots\nonumber\\
E_n(t)&=&E_n(0)+t\dot E_n(0)+\cdots
\label{six}\\
\bra{n(t)}&=&\bra{n(0)}+t\bra{\dot n(0)}+\cdots.\nonumber
\eea
Since we are interested in times less than $\eps$, which eventually will tend to zero, as long as all derivatives are finite the series will converge for sufficiently small $\eps$. 
(Note that we have not as yet assumed adiabatic evolution.)
Then
\bea
H(t)&=&\sum_n\biggl\{ \ket{n(0)}E_n(0)\bra{n(0)} + t\Bigl( \ket{n(0)}E_n(0)\bra{\dot n(0)}\nonumber\\
&&\qquad + \ket{n(0)}\dot E_n(0)\bra{n(0)}+\ket{\dot n(0)}E_n(0)\bra{n(0)}\Bigr)+O(t^2)\biggr\}
\label{seven}
\eea
In fact, as we shall see shortly, only the first term is necessary.

We can now write down an approximation of $U_0$ in \eqref{five}:
\bea
U_0&=&1-i\int_0^\eps dt \,H(t) + O(\eps^2)\nonumber\\
&=&1-i\eps\sum_n\ket{n(0)}E_n(0)\bra{n(0)} + O(\eps^2).
\label{eight}
\eea
Note that the $O(t)$ terms in \eqref{seven} give rise to $O(\eps^2)$ terms in \eqref{eight}.

We also expand $\bra{m(t_1)}$:
\beq
\bra{m(t_1)}=\bra{m(0)}+\eps\bra{\dot m(0)}+O(\eps^2).
\label{nine}
\eeq
The matrix element in \eqref{five} can now be written 
\bea
\bra{m(t_1)}U_0\ket{0(0)}&=&\left(\bra{m(0)}+\eps\bra{\dot m(0)}\right)
 \left(1-i\eps\sum_n\ket{n(0)}E_n(0)\bra{n(0}\right)\ket{0(0)}+O(\eps^2)\nonumber\\
 &=&\delta_{m0}-i\eps E_m(0)\delta_{m0}+\eps\bra{\dot m(0)}0(0)\rangle+O(\eps^2)\nonumber\\
 &=&e^{-i\eps E_m(0)}\delta_{m0}+\eps\bra{\dot m(0)}0(0)\rangle+O(\eps^2).
 \label{ten}
 \eea
 Substituting this in \eqref{five},
 \beq
 \ket{\psi(t_1)}=e^{-i\eps E_0(0)}\ket{0(t_1)}+\eps\sum_m\bra{\dot m(0)}0(0)\rangle
 \ket{m(t_1)}+O(\eps^2)
 \label{eleven}
 \eeq
 Now, we are free to choose the phase of the instantaneous eigenstates as we please; we can use this freedom to set $\bra{\dot n(t)} n(t)\rangle=0$ for all $n$,\footnote{Locally, this is true; that we must be careful globally was discovered by Berry \cite{Berry:1984jv}: the integral of this factor over a closed loop in the parameter space of the Hamiltonian is none other than Berry's phase.} so that with this choice the sum over $m$ in \eqref{eleven} does not include $m=0$.
 
 We have succeeded in writing $\ket{\psi(t_1)}$ in terms of instantaneous eigenstates of $H(t_1)$ with the coefficients given by power series in the infinitessimal parameter $\eps$. The dominant term (as long as the derivatives of $H$ are finite), not surprisingly, is $\ket{0(t_1)}$ -- not because of any assumed adiabaticity, but simply because we have not given the system enough time to have much of a chance of changing states.

We can now apply this procedure to the second time interval, and so on. We can look at this as substituting \eqref{three} into \eqref{onea}, inserting $1=\sum_{m_{j+1}}\ket{m_{j+1}(t_{j+1})}\bra{m_{j+1}(t_{j+1})}$ to the left of each $U_j$, and expanding everything in sight, rather reminiscent of derivations of the path integral form of the propagator in quantum mechanics (where, of course, position eigenstates are inserted). This results in the following fairly cumbersome expression:
\bea
\ket{\psi(T)}=\sum_{m_N}\sum_{m_{N-1}}\cdots\sum_{m_2}\sum_{m_1}\ket{m_N(T)}
 \prod_{j=0}^{N-1}\left(\delta_{m_{j+1},m_j}e^{-i\eps E_{m_j}(t_j)}
 +\eps \bra{\dot m_{j+1}(t_j)}m_j(t_j)\rangle + O(\eps^2)\right),
\label{twelve}
\eea
where $m_0$ is {\em not} summed over: $m_0=0$. For small $\eps$, in each of the factors in parentheses the first term is obviously dominant; nonetheless, in the product of these factors there are so many more contributions coming from the $\eps$ term that it is just as important when $\eps\to0$. We will show, however, that the $\eps^2$ term is indeed negligible.

Let us now dissect \eqref{twelve} by keeping track of the number of $\eps$ terms we include in the product, the remaining factors being the first term. To begin, we take the first term from every factor in the product. We obtain what will be referred to as $\ket{\psi(T)}^{(0)}$:
\beq
\ket{\psi(T)}^{(0)}=\sum_{m_N}\sum_{m_{N-1}}\cdots\sum_{m_2}\sum_{m_1}\ket{m_N(T)}
  \prod_{j=0}^{N-1}\delta_{m_{j+1},m_j}e^{-i\eps E_{m_j}(t_j)}.
 \label{thirteen}
 \eeq
 The $\delta$-functions eliminate all the sums, and in the limit $N\to\infty$ we get
 \beq 
 \ket{\psi(T)}^{(0)}=e^{-i\int_0^T dt\,E_0(t)}\ket{0(T)}.
 \label{fifteen}
 \eeq
 This term is the usual adiabatic result: the state is the instantaneous eigenstate which evolved from the initial state, multiplied by the expected dynamical phase factor.
 
 The next contribution  to $\ket{\psi(T)}$, referred to as $\ket{\psi(T)}^{(1)}$, is obtained by taking one $\eps\bra{\dot m}m\rangle$ term and otherwise taking the first term from the product in \eqref{twelve}. We must sum over which factor contributes the $\eps\bra{\dot m}m\rangle$ term (the $k$th factor, say):
\beq
\ket{\psi(T)}^{(1)}=\sum_{k=0}^{N-1}\sum_{m_N}\sum_{m_{N-1}}\cdots\sum_{m_2}\sum_{m_1}\ket{m_N(T)}
  \left(\prod_{j\neq k}\delta_{m_{j+1},m_j}e^{-i\eps E_{m_j}(t_j)}\right)
  \eps \bra{\dot m_{k+1}(t_k)}m_k(t_k)\rangle.
 \label{sixteen}
 \eeq
Only the sum over $m_{k+1}$ survives; earlier $m_k$'s are set to zero while later ones are set to $m_{k+1}$. Renaming $m_{k+1}\to m$ and $t_k\to t_1$ and taking the limit $N\to\infty$, we obtain
\beq
\ket{\psi(T)}^{(1)}=\sum_{m\neq0}\ket{m(T)}\int_0^T dt_1\,
e^{-i\int_{t_1}^T dt\, E_m(t)}\bra{\dot m(t_1)}0(t_1)\rangle e^{-i\int_{0}^{t_1} dt\, E_0(t)}.
\label{seventeen}
\eeq
This contribution represents an adiabatic evolution in the initial state (more precisely, the state evolving adiabatically from the initial state) until time $t_1$, a transition $0\to m$ at that time, followed by another adiabatic evolution in the state $\ket{m}$. So $\ket{\psi(T)}^{(1)}$ is the ``one-jump" contribution to $\ket{\psi(T)}$ (whence the superscript).

The next contribution to $\ket{\psi(T)}$, taking two $\eps\bra{\dot m}m\rangle$ terms, is the ``two-jump" contribution; the result can be written by inspection:
\bea
\ket{\psi(T)}^{(2)}&=&\sum_{m\neq n}\sum_{n\neq0}\ket{m(T)}\int_0^T dt_2\int_0^{t_2} dt_1\,
e^{-i\int_{t_2}^T dt\, E_m(t)}\bra{\dot m(t_2)}n(t_2)\rangle \nonumber\\
&&\qquad\qquad e^{-i\int_{t_1}^{t_2} dt\, E_n(t)}\bra{\dot n(t_1)}0(t_1)\rangle 
e^{-i\int_{0}^{t_1} dt\, E_0(t)}.
\label{eighteen}
\eea

The physical interpretation is clear: there are transitions at times $t_1$ and $t_2>t_1$; otherwise, the evolution is adiabatic. The generalization to higher terms is obvious.

It remains to demonstrate that the $O(\eps^2)$ terms in \eqref{twelve} are unimportant. This can be shown by straightforward power counting. For instance, if we consider taking one $O(\eps^2)$ term and otherwise taking the first term, we get an expression of the form
\beq
\eps^2\sum_k \sum_m\ket{m(T)}
\exp\left(-i\eps\sum_{j=k+1}^{N-1}E_m(t_j)\right) X_k
\exp\left(-i\eps\sum_{j=0}^{k-1}E_0(t_j)\right),
\label{nineteen}
\eeq
where $X_k$ contains various matrix elements, energies, etc. We can convert $\eps\sum_k$ and $\eps\sum_j$ to integrals, leaving one factor $\eps$ left over, indicating that this term is vanishingly small as $\eps\to0$.

Putting all the terms together, we get an expansion for $\ket{\psi(T)}$:
\beq
\ket{\psi(T)}=\ket{\psi(T)}^{(0)}+\ket{\psi(T)}^{(1)}+\ket{\psi(T)}^{(2)}+\cdots.
\label{twenty}
\eeq
We can give a diagrammatic representation for each term in the series. With time going to the right,
\beq
\ket{\psi(T)}=\ {}
\raisebox{-.5cm}{
\begin{picture}(1,1.5)
\put(0,0.5){\line(1,0){1}}
\put(.4,.6)0
\end{picture}
}
\ +\ {}
\begin{picture}(2,1)
\multiput(0,0)(1,.5){2}{\line(1,0){1}}
\put(1,0){\line(0,1){.5}}
\put(.4,.1){$0$}
\put(1.4,.6){$m$}
\put(.9,-.3){$t_1$}
\end{picture}
\ +\ {}
\begin{picture}(3,1)
\multiput(0,0)(2,.5){2}{\line(1,0){1}}
\put(1,.75){\line(1,0){1}}
\put(1,0){\line(0,1){.75}}
\put(2,0.5){\line(0,1){.25}}
\put(.4,.1){$0$}
\put(1.4,.85){$n$}
\put(2.4,.6){$m$}
\put(.9,-.3){$t_1$}
\put(1.9,-.3){$t_2$}
\end{picture}
\ +
\cdots
\label{twentyone}
\eeq
where each horizontal line represents a factor $\exp -i\int dt\, E_n(t)$, each jump $n\to m$ represents a factor $\bra{\dot m(t)}n(t)\rangle$, and we add a ket corresponding to the final state at time $T$ ($\ket{0(T)}$ in the first diagram, $\ket{m(T)}$ in the rest). Finally, the times of the jumps must be integrated over (respecting the ordering) and the states $m,n$, etc. must be summed over.

The expansion \eqref{twenty} can also be written in the following compact (albeit not terribly useful) form:
\beq
\ket{\psi(T)}=\sum_m \ket{m(T)} U_{m0}(T),
\label{twentyonea}
\eeq
where $U$ is the time evolution operator written in the ``moving" basis given by the instantaneous eigenstates. This is
\beq
U_{mn}(T)=e^{-i\int_0^T E_m(t) dt} \left( \hat T \exp X \right)_{mn},
\label{twentyoneb}
\eeq
where
\beq
X_{mn}=\int_0^T dt\,\bra{\dot m(t)}n(t)\rangle
e^{-i\int_0^T (E_n(t)-E_m(t)) dt}.
\label{twentyonec}
\eeq
In this form, our result can be seen to agree with that of \cite{Moody:1989vh}, and with \cite{Berry:1987a} for the no-transition case.

We can easily calculate the amplitude to end up in the state $\ket{m(T)}$, $\bra{m(T)}\psi(T)\rangle$; for this we can use the same diagrammatic expansion but we do not sum over the final state and do not include the final ket in the analytic expression. For example, the amplitude to end up in the state $\ket{0(T)}$ (that is, to have made no net jump) is
\bea
\bra{0(T)}\psi(T)\rangle&=&
\raisebox{-.5cm}{
\begin{picture}(1,1.5)
\put(0,0.5){\line(1,0){1}}
\put(.4,.6)0
\end{picture}
}
\ +\ {}
\begin{picture}(3,1)
\multiput(0,0)(1,.5){2}{\line(1,0){1}}
\put(2,0){\line(1,0){1}}
\multiput(1,0)(1,0){2}{\line(0,1){.5}}
\put(.4,.1){$0$}
\put(1.4,.6){$n$}
\put(.9,-.3){$t_1$}
\put(1.9,-.3){$t_2$}
\end{picture}
\ +
\cdots
\nonumber\\
&=&e^{-i\int_0^T dt\,E_0(t)}\biggl\{1+\sum_{n\neq0}\int_0^T dt_2 \int_0^{t_2} dt_1
\bra{\dot 0(t_2)}n(t_2)\rangle \nonumber\\
&&\qquad\qquad\qquad\qquad\qquad
\times e^{-i\int_{t_1}^{t_2} dt(E_n(t)-E_0(t))}
\bra{\dot n(t_1)}0(t_1)\rangle +\cdots\biggr\}
\label{twentytwo}
\eea
while the amplitude to have jumped to the state $\ket{m(T)}$ is
\bea
\bra{m(T)}\psi(T)\rangle&=&
\raisebox{-.5cm}{
\begin{picture}(2,1.5)
\multiput(0,.5)(1,.5){2}{\line(1,0){1}}
\put(1,.5){\line(0,1){.5}}
\put(.4,.6){$0$}
\put(1.4,1.1){$m$}
\put(.9,.2){$t_1$}
\end{picture}
}
\ +\ {}
\begin{picture}(3,1)
\multiput(0,0)(2,.5){2}{\line(1,0){1}}
\put(1,.75){\line(1,0){1}}
\put(1,0){\line(0,1){.75}}
\put(2,0.5){\line(0,1){.25}}
\put(.4,.1){$0$}
\put(1.4,.85){$n$}
\put(2.4,.6){$m$}
\put(.9,-.3){$t_1$}
\put(1.9,-.3){$t_2$}
\end{picture}
\ +
\cdots
\nonumber\\
&=&e^{-i\int_0^T dt\,E_0(t)}\biggl\{\int_0^{T} dt_1
e^{-i\int_{t_1}^{T} dt(E_m(t)-E_0(t))}\bra{\dot m(t_1)}0(t_1)\rangle \nonumber\\
&&\qquad\qquad\qquad+\sum_{n\neq0,m}\int_0^T dt_2 \int_0^{t_2} dt_1
e^{-i\int_{t_2}^{T} dt(E_m(t)-E_0(t))}\bra{\dot m(t_2)}n(t_2)\rangle \nonumber\\
&&\qquad\qquad\qquad\qquad\qquad \times e^{-i\int_{t_1}^{t_2} dt(E_n(t)-E_0(t))} \bra{\dot n(t_1)}0(t_1)\rangle +\cdots\biggr\}
\label{twentythree}
\eea

The expansion \eqref{twenty} for $\ket{\psi(T)}$ is obviously an expansion in powers of the number of jumps, or transitions, the system undergoes. Under what circumstances is this expansion useful; that is, under what circumstances will it converge? Intuitively, we expect this to occur when the Hamiltonian is slowly-varying since then transitions are expected to be rare. In Eqs. (\ref{fifteen},\ref{seventeen},\ref{eighteen}) the presence of increasing powers of matrix elements $\bra{\dot m} n\rangle$ suggests that indeed as the Hamiltonian varies more and more slowly, the instantaneous eigenstates will change more and more slowly, and the expansion should indeed get more and more convergent.

However, it should be noted that the smallness of the matrix elements in \eqref{twentythree} may be cancelled by the largeness of the interval over which the time is integrated. It is in fact the relatively slow variation of the matrix elements on the time scale related to the Bohr frequency that is the key to improved convergence of \eqref{twenty}, and to the validity of the adiabatic approximation.

Before examining the adiabatic case, let us look at the case of non-adiabatic evolution. A trivial observation is that for a given Hamiltonian $H(t)$, then for $T$ sufficiently small, integrals such as those in (\ref{seventeen},\ref{eighteen}) go to zero, and the system is to a good approximation in its initial state, for the simple reason that we have not given it enough time to make a transition. The important question is: How short must $T$ be for this to be so? If the scale on which the Hamiltonian varies is $\tau$, then generically $|\bra{\dot m}n\rangle|\sim(\tau)^{-1}$, and dimensional analysis indicates that $||\ket{\psi(t)}^{(k)}||\sim(T/\tau)^k$. Thus, successive terms in \eqref{twenty} are smaller if $T<\tau$, and the series will converge (at least naively; a rigorous statement about convergence would of course require a detailed analysis). We can conclude that the initial state will be depleted on a time scale $T\sim\tau$, which is as one might expect: the time scale of depletion of the initial state is none other than the time scale of the Hamiltonian.

So far, we have made no use of the other time scale available, namely, the inverse of the characteristic energy scale of the problem. This is of course the key to the adiabatic approximation. We have three time scales available: the inverse of the characteristic energy scale ($E$, say) of the problem, the rate of change of the Hamiltonian $\tau$, and the duration of the process $T$. It is common to consider the latter two time scales to be equal, but for generality we will not do so here.

To more easily keep track of the relative size of the various terms, it is useful to make a change of notation to dimensionless quantities. The Hamiltonian is characterized by a time scale $\tau$ and an energy scale $E$, so we define a dimensionless time $s=t/\tau$, and a dimensionless Hamiltonian and its energies and eigenstates which depend smoothly upon $s$; these are $h(s)=H(t)/E$, $\epsilon_n(s)=E_n(s)/E$ and $\ket{\hat n(s)}=\ket{n(t)}$. We make the following assumptions:
\begin{enumerate}
\item The state $\ket{\hat 0(s)}$ has zero energy, $\eps_0(s)=0$. (This assumption is for simplicity only and can easily be removed with only minor modifications to the analysis.)
\item The minimum energy gap to any other state is unity: $|\eps_n(s)|\ge1\ \forall\ n\neq0$.
\item Otherwise, $h(s)$ is ``generic": derivatives of the energies $\eps_n(s)$ are of order 1 (for $n\neq0$, of course), and derivatives of the states $\ket{\hat n(s)}$ have norm of order 1. (This is simply the statement that $H(t)$ is characterized by time scale $\tau$, recast in terms of dimensionless quantities.)
\end{enumerate}

The duration of the process is $T$, so we define a dimensionless duration $S=T/\tau$. We define a parameter $\la$ by $\la=E\tau$; when $\la\gg1$, the Hamiltonian can be said to evolve slowly, and the evolution of the system is expected to be adiabatic. We also define a parameter $\gamma$ by $\gamma=\max_{s;m\neq0}
|\bra{{\hat m}'(s)}\hat 0(s)\rangle|$, where the prime denotes differentiation with respect to the argument; $\gamma$ is of order unity. 

We can now rewrite the expansion \eqref{twenty} in terms of dimensionless quantities. For instance, the first correction is, from \eqref{seventeen},
\beq
|\hat\psi(S)\rangle^{(1)}=\sum_{m\neq0}\ket{\hat m(S)}\int_0^S ds_1\,
e^{-i\la\int_{s_1}^S ds\, \eps_m(s)}\bra{{\hat m}'(s_1)}\hat 0(s_1)\rangle.
\label{twentyfour}
\eeq
Since the zeroth term $|\hat\psi(S)\rangle^{(0)}$ is of norm unity, a first indication of the convergence of the series, and the validity of the adiabatic approximation, comes from the norm squared of the first-order correction:
\beq
^{(1)}\hspace{-1pt}\langle\hat\psi(S)|\hat\psi(S)\rangle^{(1)}
 =\sum_{m\neq0}\left|A_m^{(1)}(S)\right|^2,
\label{twentyfive}
\eeq
where
\beq
A_m^{(1)}(S)=\langle\hat m(S)|\hat\psi(S)\rangle^{(1)}=\int_0^S ds_1\,
e^{-i\la\int_{s_1}^S ds\, \eps_m(s)}\bra{{\hat m}'(s_1)}\hat 0(s_1)\rangle.
\label{twentysix}
\eeq
Note that the smallness of $|A_m^{(1)}|$ is not due to the smallness of the matrix element (which is of order one). To see that $|A_m^{(1)}|$ goes to zero in the limit $\la\to\infty$, we note that since $\eps_m$ and its derivatives are of order 1, the exponential oscillates rapidly, on a (dimensionless) time scale $(\la)^{-1}$. The matrix element, in contrast, varies on time scale 1, so that within one period of the exponential the matrix element is approximately constant, and the exponential integrates to zero. Corrections to this give a result of order $1/\la$.

As a special case, if $\eps_m(s)$ and the matrix element are both constants (equal to one, say), then
\beq
|A_m^{(1)}|={\left| 1-e^{-i\la S}\right| \over \la},
\label{twentyseven}
\eeq
which clearly goes to zero as $1/\la$.

To analyze $A_m^{(1)}$ in the general case, we define $f_m(s)=\int_0^s ds' \eps_m(s')$ and $g_m(s)=\bra{{\hat m}'(s)}\hat 0(s)\rangle$. Then
\beq
A_m^{(1)}=e^{-i\la f_m(S)}\int_0^S ds\, e^{i\la f_m(s)}g_m(s).
\label{twentyeight}
\eeq
The assumptions stated above tell us that $f_m$, $g_m$ and their derivatives are of order 1, and that $|f_m'|\ge1$. It is then easy to show that if $S\sim1$ and $\la\gg1$,
\beq
A_m^{(1)}={1\over i\la}\left( {g_m(S)\over \eps_m(S)}
-e^{-i\la f_m(S)}{g_m(0)\over \eps_m(0)} \right) + O(\la^{-2}).
\label{thirty}
\eeq
This clearly demonstrates that the dominant contribution to $|A_m^{(1)}|$ is the first term when $\la\gg1$; rewriting it in terms of dimensionful quantities,
\beq
A_m^{(1)}= {\bra{{\dot m}(T)} 0(T)\rangle\over i E_m(T)}
-e^{-i\int_0^T dt\, E_m(t)}{\bra{{\dot m}(0)}0(0)\rangle\over i E_m(0)} + O((ET)^{-2}).
\label{thirtyone}
\eeq
Interestingly, the dominant correction depends only on the initial and final values of the matrix element; in particular, in the special case where initially and finally the instantaneous eigenstates are constant, the dominant correction is zero and the adiabatic approximation should be more robust, being valid up to corrections of order $1/(ET)^2$ rather than $1/(ET)$.

Assuming this special case does not apply to the system under investigation, we can now give a bound for $|A_m^{(1)}|$ in terms of quantities defined earlier. The matrix elements satisfy $|g_m|\ge\ga$, while the dimensionless energies $\eps_m$ are at least unity. So
\beq
|A_m^{(1)}|\leq {2\ga\over\la}={2\max |\bra{\dot m(t)}0(t)\rangle|\over E}.
\label{thirtytwo}
\eeq
Thus, we would conclude on the basis of this first-order analysis that the adiabatic approximation is valid if the right hand side of \eqref{thirtytwo} is small; this is essentially the ``standard" condition of validity of the adiabatic approximation which has been re-examined recently \cite{Marzlin:2004, Sarandy:2004, Pati:2004, Tong:2004, Ambainis:2004, Wu:2004, Tong:2005, Ye:2005}. This leads us to the obvious question: Does the first-order analysis extend to higher orders; that is to say, is $|A_m^{(k)}|\sim(E\tau)^{-k}$? As we will see, the answer is no. In a way, this is not so surprising: if indeed $A_m^{(k)}$ behaved in this way, we would conclude that a slowly-evolving Hamiltonian will {\em never} give rise to a transition with any appreciable probability. Surely this is not so; for fixed rate of change of the Hamiltonian, the system ought to {\em eventually} escape from the instantaneous eigenstate related to the initial state.

The fundamental reason why the first-order analysis does not extend directly to higher orders is most easily seen by looking at $A_0^{(2)}$, the second term in \eqref{twentytwo}. We can write the dimensionless version of this as
\beq
A_0^{(2)}(S)=\sum_n\int_0^S ds_2\langle \hat 0'(s_2)|\hat n(s_2)\rangle A_n^{(1)}(s_2).
\label{thirtythree}
\eeq
From \eqref{thirty}, we see that $A_n^{(1)}$ contains a non-oscillating piece, and the argument used to show the smallness of \eqref{twentysix} does not apply to \eqref{thirtythree}; indeed,
\beq
A_0^{(2)}(S)\sim {S\over\la}={T\over\tau}{1\over E\tau}.
\label{thityfour}
\eeq
Rather than being a factor $(E\tau)^{-1}$ times the first-order term, it is a factor of order $T/\tau$ times the first-order term. Taken at face value, we would conclude that the expansion no longer converges, and presumably the initial state is depleted, on a time scale $T\sim\tau$: we would be no better off than the non-adiabatic evolution. However, the situation is not quite as dire as this: higher-order terms can be shown to get smaller up until a time of order $T\sim E\tau^2$. Thus, the depletion time scale appears to be longer than the non-adiabatic case by a factor $E\tau\gg 1$ (the inequality being true for a slowly-varying Hamiltonian), so indeed some time is gained in the adiabatic limit. This will be analyzed in more detail in a future publication \cite{MacKenzie:inprogress}.

As a final remark, we suspect that one factor giving rise to the apparent inadequacy of the ``standard" condition of validity of the adiabatic approximation lies in the fact that the examples given in some papers questioning this condition may not fit into the type of problem discussed in this paper (and in at least some of the standard discussions of the adiabatic method, such as that of Messiah \cite{Messiah:1962}), that is to say, they are not described by a Hamiltonian which can be written $H(t)=E h(t/\tau)$, where $h(s)$ is ``generic" as defined above. More precisely, a Hamiltonian containing a part which is rapidly oscillating but small in amplitude can satisfy the ``standard" condition, yet the evolution might not necessarily be adiabatic; such a Hamiltonian is not ``generic" as defined above. This is currently under investigation \cite{MacKenzie:inprogress}.

Note added: After this paper was submitted, a preprint by T. V\'ertesi and R. Englman has appeared, which studies possible failures of the standard condition of validity of the adiabatic approximation from a perturbative point of view \cite{Vertesi:2005}.
\bigskip

We thank A.A. M\'ethot, M.B. Paranjape, and B. Sanders for useful conversations.
This work was funded in part by the
National Science and Engineering Research Council.


\bibliographystyle{unsrt}
\bibliography{fieldtheory}


\end{document}